\documentclass[referee]{raa}                  

\usepackage{graphicx,times}             
\input{epsf.sty}                        
\input{psfig.sty}                       

\def\Mesz{M\'esz\'aros~}

\def\etal{et al.}
\def\beq{ \begin{equation} }
  \def\eeq{ \end{equation} }
\def\beqa{ \begin{eqnarray} }
  \def\eeqa{ \end{eqnarray}   }

\begin{document}

\title{Early afterglows from radially
  structured outflows and the application to X-ray shallow
  decays
}

\volnopage{Vol.0 (200x) No.0, 000--000}  
\setcounter{page}{1}                     

\author
{
  Xue-Wen Liu\mailto{}
  \inst{1,2,3}
  \and Xue-Feng Wu\mailto{}
  \inst{1,3,4}
  \and Yuan-Chuan Zou\mailto{}
  \inst{5}
  \and Tan Lu\mailto{}
  \inst{1,3}
}

\institute{Purple Mountain Observatory, Chinese Academy of
  Sciences, Nanjing, 210008, China\\
  \email{astrolxw@gmail.com,~xfwu@pmo.ac.cn,~yuanchuan@gmail.com,~t.lu@pmo.ac.cn}
  \and
  Graduate School, Chinese Academy of Sciences, Beijing, 100039, China\\
  \and
  Joint Center for Particle Nuclear Physics and Cosmology
  (J-CPNPC), Nanjing 210093, China\\
  \and
  Department of Astronomy \& Astrophysics, Pennsylvania State
  University, University Park, PA 16802, USA\\
  \and
  Department of Physics, Huazhong University of
  Science and Technology, Wuhan 430074, China
}

\date{Received~~; accepted~~}

\abstract
{
  In the fireball model, it is more physically realistic
  that gamma-ray burst (GRB) ejecta have a range of bulk Lorentz
  factors (assuming $M\propto \Gamma^{-s}$). The low Lorentz factor part of the ejecta
  will catch up with the high Lorentz factor part when the latter
  is decelerated by the surrounding medium to a
  comparable Lorentz factor. Such a process will develop a long-lasting weak
  reverse shock until the whole ejecta are
  shocked. Meanwhile, the forward shocked materials are gradually
  supplied with energy from the ejecta that are catching-up, and thus the temporal decay of the
  forward shock emission will be slower than that without an
  energy supply. However, the reverse shock may
  be strong. Here, we extend the standard reverse-forward
  shock model to the case of radially nonuniform ejecta.
  We show that this process can be classified into
  two cases: the thick shell case and the thin shell case. In the
  thin shell case, the reverse shock is weak and the
  temporal scaling law of the afterglow is the same as that in
  Sari \& \Mesz (2000). However, in the thick shell case, the reverse
  shock is strong and thus its emission dominates the afterglow
  in the high energy band. Our results also show 
  slower decaying behavior of the afterglow due to the energy
  supply by low Lorentz factor materials, which may
  help the understanding of the plateau observed in the early
  optical and X-ray afterglows.
}

\authorrunning{Xue-Wen Liu, X. F. Wu, Y. C. Zou \& T. Lu }

\maketitle

%
%
\section{Introduction}        
\label{sect:intro} The central engine and surrounding environment
provide the most important insights to the mystery of gamma-ray
bursts (GRBs), the most violent explosions in the universe. Thanks
to ${\textit BeppoSAX}$, owing to its ability to accurately
locate objects, the first afterglow of a GRB was discovered in 1997
(Costa et al. 1997). Afterward, broad band data of afterglows were
achieved, and were fitted using the standard
fireball-shock model (Rees \& \Mesz 1992, 1994; \Mesz \& Rees
1993, 1997; for reviews, see Zhang 2007). The parameters of
GRBs, such as the total burst energy, the type (ISM or wind) and
number density of the environment, and the electron and magnetic field
equipartition factors were then constrained (e.g., Wu et al. 2003;
Fan et al. 2002; Zhang et al. 2003). Early afterglows can even be
used to constrain the initial Lorentz factors of GRB fireballs
(Molinari et al. 2007; Xue et al. 2009). It is believed that early
afterglows are produced by reverse-forward shocks when 
relativistic ejecta interact with the circum-burst medium, which
was first studied by Rees \& \Mesz (1992) and Sari \& Piran
(1995). Then GRB 990123, a remarkable event with a bright early
optical flash was discovered (Akerlof et al. 1999), which
was interpreted by the reverse-forward external shock model well(Sari \&
Piran 1999). Much attention was consequently focused on early
afterglow radiation, considering the effects of circum-burst
environments and non-relativistic reverse shocks (Kobayashi 2000;
Wu et al. 2003; Zou et al. 2005). Subsequently, more optical
flashes were observed in, e.g., GRBs 021211, 050525a, 060111B,
060117B and 080319B (Wei 2003; Shao \& Dai 2005; Klotz et al.
2006; Jelinek et al. 2006; Racusin et al. 2008). These works were
based on the assumption that the Lorentz factor does not change in
the shell. The whole light curve from the reverse-forward external
shock has two types: re-brightening (Type I) and flattening (Type
II) (Zhang et al. 2003). In some GRBs, such as GRBs 050319,
060206, 060210 and 060313, the early optical light curves have a
plateau which is difficult explain within the uniform
ejecta model (UEM). Meanwhile, the shallow decay of the
canonical X-ray afterglow behavior discovered in the
$\textit{Swift}$ era (Nousek 2006; Zhang et al. 2006; O'Brien et
al. 2006) remains a matter of debate.

Because of the above problems, we reconsider the
baryon-dominated energy injection model in which the ejecta have a
wide $\rm {\Gamma}$-distribution: the part of the ejecta with low
Lorentz factor lagging behind the one with high Lorentz factor.
The low-$\Gamma$ part catches up with the high-$\Gamma$ part when
the latter is decelerated to a comparable Lorentz factor, so the
reverse shock is usually mildly relativistic and mainly contributes
to the far-IR or millimeter band (Rees \& \Mesz 1998; Sari \Mesz
2000). Once the reverse shock starts, it will travel through the
whole ejecta from the front highest-$\Gamma$ part to the rear
lowest-$\Gamma$ part. Based on the treatment widely adopted
in the UEM, we reconsider this issue by assuming a given distribution of
Lorentz factors in the ejecta. We calculate the dynamic evolution
of the reverse-forward shocks produced by this radially structured
ejecta propagating into the circum-burst medium, and present the
analytical and numerical results.

We organize our paper as follows. In $\S 2$, we describe the
dynamics of the reverse-forward shocks including the thick shell
case and the thin shell case, respectively. We discuss the reverse
shock emission in $\S 3$. The numerical results are shown in $\S
4$. Finally, we present a brief discussion in $\S 5$.

\section{DYNAMICS OF THE EJECTA WITH A $\Gamma$-DISTRIBUTION}
\label{sec:dyn} 
\begin{figure}
\centering
\includegraphics[angle=360,width=1.0\textwidth]{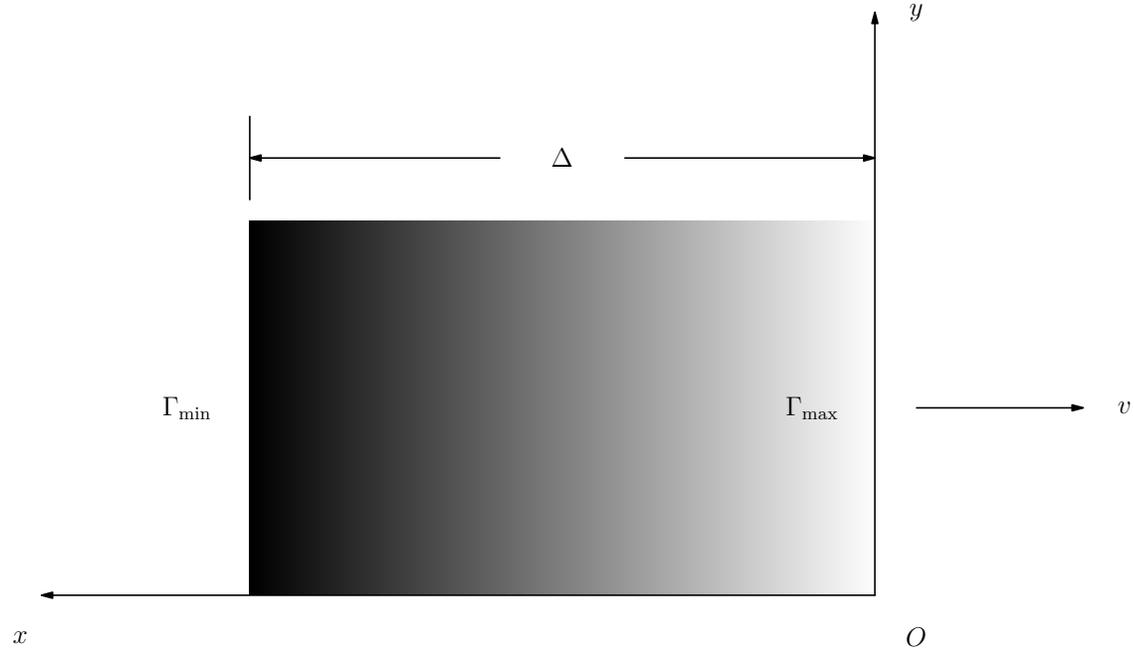}\\
\caption{Schematic of the
  radially structured shell. $\Delta$ is the width of the shell,
  $\Gamma_{\rm min}$ and $\Gamma_{\rm max}$ are the Lorentz factors
  on the edge of the shell respectively. The part with the lower
  Lorentz factor in the shell has more kinetic energy
  (marked with darker gray) than that with the higher Lorentz factor.}
\label{fig1}
\end{figure}

As Rees \& \Mesz (1998) postulated, the central
engine of GRBs may eject relativistic shell-like ejecta  with a
range of Lorentz factors
\beq
M(>\Gamma) \propto \Gamma^{-s}.
\eeq
Such ejecta have an energy $\Gamma Mc^2 \propto \Gamma^{-s+1}$.
Applying the model to observations shows that the value of
the index $s$ is typically $\sim 2.5$ (Zhang et al. 2006; Nousek
et al. 2006), which is larger than the suggested value $\sim1.5$
(e.g., Rees \& \Mesz 1992), so the low-$\Gamma$ mass carries more
kinetic energy than predicted. When the ejecta interact with the
circum-burst medium, a pair of shocks emerges: a forward shock
propagating into the circum-burst medium and a reverse shock
propagating into the shell. There are four regions separated by
the two shocks: (1) the unshocked circum-burst medium, (2) the
shocked medium, (3) the shocked shell material and (4) the
unshocked shell material. Using the shock jump condition and the
equality of pressure and velocity along the contact discontinuity,
the Lorentz factor $\Gamma$ and the number density in both shocked
media can be determined by the density of the circum-burst medium
$n_1$ and the unshocked shell $n_4$ (Blandford \& McKee 1976,
hereafter BM). Here, the number density of the unshocked shell is
nonuniform, depending on the Lorentz factor distribution in the
shell
\beq
n_4=\frac{|(dM(>\Gamma)/d\Gamma)(d\Gamma/dx)|}{4\pi r^2 m_p \Gamma(x)}.
\eeq
We assume an initial Lorentz factor distribution in the shell
\beq\label{eq:Gdis}
\Gamma\sim \Gamma_{\rm min}\left[\frac{x(t=0,\Gamma)}
  {\Delta_0}\right]^{-1/b},
\eeq where $\Delta_0$ and $\Gamma_{\rm min}$ are the initial width
and minimum Lorentz factor of the shell, $x(t=0,\Gamma)$
represents the initial position in the shell with the origin
located at the outer edge of the shell (See Fig. \ref{fig1}). Due
to the distribution of Lorentz factors, the shell will spread with
time, then the value of $x$ of a fixed element will become larger
and larger,
\beq\label{eq:coordinate}
x(t,\Gamma)=x(t=0,\Gamma)+(\beta_{\rm max}-\beta)ct ,
\eeq
where $\beta=\sqrt{1-1/\Gamma^2}$. Thus the Lorentz factor
distribution in the shell at any time is determined.

The properties of the shocks are largely determined by the
parameter defined as \beq\label{eq:f}
f\equiv\frac{n_4}{n_1}=\frac{sM_b\Gamma_{\rm min}^s\Gamma^{-s-2}}
{4\pi r^2 m_p n_1}\Big|\frac{d\Gamma}{dx}\Big| , \eeq where $M_b$
is the total mass of the shell. Eq. (1) can be written as
$M(>\Gamma)\approx M_b\Gamma_{\rm min}^s\Gamma^{-s}$ (here
$\Gamma_{\rm min} \ll \Gamma_{\rm max}$ is assumed). 
Combining Eq. (\ref{eq:coordinate}) and (\ref{eq:f}), we
need another equation to describe the evolution of the radius and
Lorentz factor of the shocks: the relation between the distance
$dx$ which the reverse shock travels in the shell and the distance
$dr$ which the shell propagates in the circum-burst medium in the
same time interval is (Kobayashi 2000)
\beq\label{eq:drdx}
dr=\alpha \Gamma f^{1/2}dx ,
\eeq
where $r$ is the radius of the shell and the parameter
$\alpha$ is $\sim 1$.

There are two approximations under which the shock evolution can
be described analytically: the thick shell case and the thin shell
case, depending on the significance of the spreading effect in the
last term in Eq. (\ref{eq:coordinate}). If the spreading term is
larger than the initial $x(t=0,\Gamma)$, the shell is regarded as
a thin shell, otherwise it is a thick shell. Below, we will
consider these two cases separately.

\subsection{Thick Shell Case}
In the thick shell case,  the spreading effect can be
ignored, so the width of the shell always remains at its
initial value $\Delta_0$ during the time that the reverse
shock is crossing the shell.

We can get from Eqs. (\ref{eq:Gdis}) and (\ref{eq:coordinate})
\beq dx=-b\Delta_0\Big(\frac{\Gamma}{\Gamma_{\rm
min}}\Big)^{-b}\frac{d\Gamma}{\Gamma}. \eeq 
We can now calculate
the comoving number density $n_4$ and the density ratio $f$.
According to Eq. (\ref{eq:f}), we have 
\beq\label{eq:thickG}
\frac{dr}{dx}=\frac{\alpha^2sM_b\Gamma_{\rm
    min}^s\Gamma^{-s}}
{4\pi r^2 m_p n_1}\Big|\frac{d\Gamma}{dr}\Big| . \eeq 
In general,
the number density of the circum-burst medium can be modeled as
$n_1=Ar^{-k}$. Specifically, $A=n_1=1.0~n_{1,0}~\rm cm^{-3}$ for
an ISM environment ($k=0$) and $A=3 \times 10^{35}~A_{\ast}~\rm
cm^{-1}$ for a free wind environment ($k=2$) (Chevalier \& Li
2000). Throughout this work, we adopt the convention $Q_x=Q/10^x$
in cgs units. The solution of Eq. (\ref{eq:thickG}) reads \beq
r^{4-k}=\Big(\frac{4-k}{s+b-1}\Big)^2\frac{b(s-1)E\Delta_0\alpha^2}
{4\pi Am_pc^2} \Big(\frac{\Gamma}{\Gamma_{\rm
min}}\Big)^{-(s+b-1)} . \eeq Keep in mind that $dr=2\Gamma^2cdt$,
the evolution of the Lorentz factor and radius of the ejecta with
time can now be described. Just when the reverse shock crosses the
shell, the Lorentz factor is equal to $\Gamma_{\rm min}$ and the
shell reaches the crossing radius
\begin{eqnarray}
r_\Delta&=&\Big[\Big(\frac{4-k}{s+b-1}\Big)^2\frac{b(s-1)E\Delta_0
\alpha^2}{4\pi Am_pc^2}\Big]^{\frac{1}{4-k}}\nonumber \\
&=& \cases{ 9.4\times 10^{16}\alpha^{1/2}\Big[\frac{b(s-1)}{(s+b-1)^2}
\Big]^{1/4}E_{53}^{1/4}\Delta_{0,12}^{1/4}n_{1,0}^{-1/4}\rm cm, &$(k=0)$ \cr
8.5\times 10^{15}\alpha\Big[\frac{b(s-1)}{(s+b-1)^2}\Big]^{1/2}
E_{53}^{1/2}\Delta_{0,12}^{1/2}A_{\ast}^{-1/2}\rm cm,
&$(k=2)$} .
\end{eqnarray}
In the thick shell case, the spreading effect is always negligible
which requires $r_\Delta \leq \Gamma_{\rm min}^2\Delta_0$.
Thus a lower limit of the minimal Lorentz factor must be satisfied
\beq \Gamma_{\rm min}\geq
\cases{310\Big[\frac{b(s-1)}{(s+b-1)^2}\Big]^{1/8}
\alpha^{1/4}E_{53}^{1/8}\Delta_{0,12}^{-3/8}n_{1,0}^{-1/8},
~~$(\textit{k}=0)$ \cr
92\Big[\frac{b(s-1)}{(s+b-1)^2}\Big]^{1/4}\alpha^{1/2}E_{53}^{1/4}
\Delta_{0,12}^{-1/4}A_{\ast}^{-1/4},
~~$(\textit{k}=2)$}. \eeq
Whether or not the reverse shock is relativistic depends on
the parameter
\beq \frac{f}{\Gamma^2}\Big|_{\Gamma_{\rm min}}\approx
\cases{5.75\frac{(s+b-1)\sqrt{s-1}}{b^{3/2}}\alpha^{-1}E_{53}^{1/2}n_{1,0}^{-1/2}
\Delta_{0,12}^{-3/2}\Gamma_{\rm min,2}^{-4},~~$(\textit{k}=0)$\cr
0.17\frac{s-1}{b}E_{53}A_\ast^{-1}\Delta_{0,12}^{-1}\Gamma_{\rm
min,2}^{-4},~~~~~~~~~~~~~~~~~~~~~~~~~~~~~~$(\textit{k}=2)$}.
\eeq
It can be seen that the parameter $f/\Gamma^2$ is smaller than one
for the typical wind case (k=2) and for a dense ISM case ($n_1=100$) at
the crossing time, and before the crossing time it is proportional
to $\Gamma^{2(3b-bk+2k-s-7)/(4-k)}$. If the Lorentz factor
distribution is not too steep $0< b < 3$ and the mass distribution
index has a typical value of $1.5\leq s \leq 2.5$, $f<\Gamma^2$ holds
for the entire reverse-forward shocks interaction period which
means that the reverse shock is relativistic all along. While in
the other parameter space, it is possible that $f\gg\Gamma^2$ at the
initial stage when the reverse shock is non-relativistic, and then
$f\leq\Gamma^2$ which corresponds to the reverse shock evolving
from being non-relativistic to being relativistic. If the reverse
shock is relativistic, the relative Lorentz factor $\gamma_{34}$
between the shocked shell and the unshocked shell is
$\sim(\Gamma/2)^{1/2}/f^{1/4}\gg 1$. We derive the analytical
solution for the former in Section 3.1.

\subsection{Thin Shell Case}
In the thin shell case, the spreading effect is dominant. The
position of an element in the shell, $x$, can be approximated by
\beq
x(\Gamma)\approx(\beta_{\rm
  max}-\beta)ct\approx\frac{r}{2\Gamma^2} .
\eeq
the parameter $f$ can be written as \beq\label{eq:fthin}
f=\frac{(s-1)E}{4\pi r^3m_pc^2n_1} \Big(\frac{\Gamma}{\Gamma_{\rm
min}}\Big) ^{-(s-1)} . \eeq Based on the same procedure applied in
the thick shell case, we can obtain the relation between the
radius and the Lorentz factor
\beq\label{eq:rthin}
r^{3-k}=\Big(\frac{3-k}{s+1}\Big)^2\frac{\alpha^2(s-1)E}{4\pi
Am_pc^2\Gamma_{\rm min}^2}\Big(\frac{\Gamma}{\Gamma_{\rm min}}\Big)^{-(s+1)}.
\eeq
Then the crossing radius is
\begin{eqnarray}
r_\Delta&=&\Big[\Big(\frac{3-k}{s+1}\Big)^2\frac{\alpha^2(s-1)E}
{4\pi Am_pc^2\Gamma_{\rm min}^2}\Big]^{\frac{1}{3-k}}\nonumber \\
&=& \cases{ 1.7\times 10^{17}\Big[\frac{\alpha^2(s-1)}{(s+1)^2}
\Big]^{1/3}E_{53}^{1/3}n_{1,0}^{-1/3}\Gamma_{\rm min,2}^{-2/3}\:\rm
cm ,
\:&$(k=0)$ \cr
1.8\times 10^{15}\Big[\frac{\alpha^2(s-1)}{(s+1)^2}
\Big]E_{53}A_{\ast}^{-1}\Gamma_{\rm min,2}^{-2}\:\rm cm,\:
&$(k=2)$} .
\end{eqnarray}
In contrast to the thick shell case, an upper limit of the minimal
Lorentz factor must be satisfied to keep the thin shell assumption
valid all along
\beq
\Gamma_{\rm min}\leq
\cases{290\Big[\frac{\alpha^2(s-1)}{(s+1)^2}
\Big]^{1/8}
E_{53}^{1/8}\Delta_{0,12}^{-3/8}n_{1,0}^{-1/8}, &$(k=0)$ \cr
65\Big[\frac{\alpha^2(s-1)}{(s+1)^2}\Big]^{1/4}E_{53}^{1/4}
\Delta_{0,12}^{-1/4}A_{\ast}^{-1/4}, &$(k=2)$} .
\eeq
For the thin shell case, it is interesting that
$f/\Gamma^2\equiv\alpha^{-2}({s+1}/{3-k})^2$, which means that the
reverse shock is always mildly relativistic ($\gamma_{34}-1\approx
\Gamma^2/f \sim 1$).

\section{REVERSE SHOCK EMISSION}
Now that the dynamic related parameters, i.e., $\Gamma$, $\gamma_{34}$
and $n_4$ are determined, the radiation related properties of
the shocked materials, such as the strength of the magnetic field
$B_i'$, the minimum Lorentz factor $\gamma_{m,i}'$, the cooling
Lorentz factor $\gamma_{c,i}'$ and the number $N_{e,i}$ of shocked
electrons can be determined. For the shocked region,
the fraction $\epsilon_B$ and $\epsilon_e$ of the internal energy
are assumed to be carried by magnetic fields and shock-accelerated
electrons, respectively. The co-moving magnetic field is equal to
$\sqrt{8\pi\epsilon_Be_i'}$, where the internal energy density
$e_i'=(\gamma_{\rm rel}-1)m_pc^2n_{i}$ (for the forward shock
$\gamma_{\rm
  rel}\equiv\gamma_{2}\approx1/2\Gamma^{1/2}f^{1/4}$; for the
reverse shock $\gamma_{\rm rel}\equiv\gamma_{34}$). The minimum
Lorentz factor $\gamma_{m,i}'=\epsilon_eC_p(m_p/m_e)(\gamma_{\rm
rel}-1)$ with $C_p\equiv(p-2)/(p-1)$, the cooling Lorentz factor
$\gamma_{c,i}'=6\pi m_ec/(\sigma_TB_i'^2\gamma_i t)$. The increase
of the number of shocked electrons is $dN_{e,3}=4\pi r^2\Gamma
n_4dx$ for the reverse shocked region and $dN_{e,2}=4\pi r^2n_1dr$
for the forward shocked region. In the standard synchrotron
radiation model, the two characteristic frequencies and the peak
flux density are \beq\label{eq:radch}
\nu_{m,i}=\frac{q_eB_i'}{2\pi m_ec}\gamma_{m,i}'^2\gamma_i,
~~\nu_{c,i}=\frac{q_eB_i'}{2\pi m_ec}\gamma_{c,i}'^2\gamma_i,
~~F_{\nu,\rm max,i}=\frac{N_{e,i}}{4\pi
d_L^2}\frac{m_ec^2\sigma_T}{3q_e}B_i'\gamma_i, \eeq where $d_L$ is
the luminosity distance of a GRB, $q_e$ is the charge of electron
and $\sigma_T$ is the Thomson cross section. The temporal indices
of these two frequencies and the peak flux density as a function
of time are listed in Table 1 for both the forward shock and
reverse shock, and for both the thick shell case and thin shell
case.

The distinct discrepancy between the nonuniform ejecta model (NUEM)
and the UEM is the reverse shock emission. Once the reverse shock has
crossed the shell, the forward shock and shocked region begin to
approach the Blandford-McKee (BM) solution (Kobayashi et al.
1999), and the following light curve is the same as that in the UEM.
Below we only discuss the synchrotron emission from the
reverse-shocked region before the crossing time. Since we have
already obtained the temporal indices of $\nu_m$, $\nu_c$ and
$F_{\rm max}$, in the following we only need to know the values of
the characteristic frequencies and the peak flux density at the
crossing time so we can extrapolate the early light
curve back in time from the reverse shock.

\subsection{Thick Shell Case}
The reverse shock in the thick shell case we consider here is
assumed to be relativistic ($\gamma_{34}-1\approx \gamma_{34}$).
The crossing time $T_\Delta$ is $\sim \Delta/c$, when the Lorentz
factor of the shell is $\Gamma_{\rm min}$ and the number of the
shocked electrons is the total number of electrons in the ejecta,
i.e., $N_{e,3}=E(s-1)/(sm_pc^2\Gamma_{\rm min})$. According to Eq.
(\ref{eq:radch}), we have \beq \nu_m\sim 4.0\times
10^{13}(1+z)^{-1}\Big(\frac{p-2}{p-1} \Big)^2\epsilon_{e,-0.5}^2
\epsilon_{B,-2}^{1/2} n_{1,0}^{1/2}\Gamma_{\rm min,2.5}^2 \: \rm
Hz , \eeq

\beq
\nu_c\sim 1.0\times 10^{17}(1+z)^{-1/2}
\sqrt{\frac{\alpha^2b^3}{(s+b-1)^2(s-1)}}
\epsilon_{B,-2}^{-3/2}E_{53}^{-1/2}
n_{1,0}^{-1}T_{\Delta,-2}^{-1/2} \: \rm Hz ,
\eeq

\beq F_{\nu,\rm max}\sim 1.36(1+z)^{7/4}
\Big[\frac{(s-1)^5(s+b-1)^2}{\alpha^2b^3s^4}\Big]^{1/4}
D_{28}^{-2}\epsilon_{B,-2}^{1/2}E_{53}^{5/4}n_{1,0}^{1/4}
\Gamma_{\rm min,2.5}^{-1}T_{\Delta,2}^{-3/4}\: \rm Jy , \eeq for
the ISM case, and \beq \nu_m\sim 4.5\times
10^{14}(1+z)^{-1/2}\Big(\frac{p-2}{p-1}\Big)^2
\sqrt{\frac{(s+b-1)^2}{\alpha^2b(s-1)}}
\epsilon_{e,-0.5}^2\epsilon_{B,-2}^{1/2}
E_{53}^{-1/2}A_{\ast,-0.5}\Gamma_{\rm
  min,2.5}^2T_{\Delta,2}^{-1/2} \: \rm Hz ,
\eeq

\beq
\nu_c\sim 4.0\times 10^{14}(1+z)^{-3/2}
\sqrt{\frac{\alpha^6b^5(s-1)}{(s+b-1)^6}}
\epsilon_{B,-2}^{-3/2}E_{53}^{1/2}
A_{\ast,-0.5}^{-2}T_{\Delta,2}^{1/2} \: \rm Hz ,
\eeq

\beq F_{\nu,\rm max}\sim 6.5(1+z)^2\Big[\frac{(s+b-1)(s-1)}{\alpha
bs}\Big] D_{28}^{-2}\epsilon_{B,-2}^{1/2}E_{53}A_{\ast,-0.5}^{1/2}
\Gamma_{\rm min,2.5}^{-1}T_{\Delta,2}^{-1}\: \rm Jy , \eeq for the
wind case, where $T_{\Delta,2}=T/100s$. The above expressions for
the synchrotron radiation at the crossing time are quite similar
to those in the UEM (e.g., Kobayashi 2000; Wu et al. 2003). We find
that for a set of combinations of reasonable parameter values
($s=2.5, 2, 1.5$ and $b=2,1$) the $t_{\rm cm}$ (the time when
$\nu_m=\nu_c$) is always small for the ISM case, indicating that
the reverse-shocked electrons are always in the slow cooling
region. However, for the wind case, the electrons are usually fast
cooling during the entire reverse shock phase, because $\nu_m$ is
typically larger than $\nu_c$ at the crossing time and $\nu_m$
decreases with time while $\nu_c$ increases with time before the
crossing time.

\subsection{Thin Shell Case}
In the thin shell case, the reverse shock is always
mildly-relativistic ($\gamma_{34}-1\sim 1$) and the crossing time
depends on the crossing radius as $t_{\Delta}\propto
r_{\Delta}^{(7+s-2k)/(s+1)}$. For convenience, we choose $s=2$ to
give the typical values of the two characteristic frequencies and
peak flux density of synchrotron radiation, \beq \nu_m\sim
2.1\times 10^{12}(1+z)^{-1}\Big(\frac{p-2}{p-1}\Big)^2\alpha^4
\epsilon_{e,-0.5}^2\epsilon_{B,-2}^{1/2} \Gamma_{\rm min,1.8}^2 \:
\rm Hz , \eeq

\beq
\nu_c\sim 7.1\times 10^{16}(1+z)^{-1}\alpha^{-4/3}
\epsilon_{B,-2}^{-3/2}E_{53}^{-2/3}n_{1,0}^{-5/6}
\Gamma_{\rm min,1.8}^{4/3} \: \rm Hz ,
\eeq

\beq F_{\nu,\rm max}\sim
2.3(1+z)D_{28}^{-2}\epsilon_{B,-2}^{1/2}E_{53} n_{1,0}^{1/2}
\Gamma_{\rm min,1.8}\: \rm Jy , \eeq for the ISM case, and \beq
\nu_m\sim 4.2\times
10^{12}(1+z)^{-1}\Big(\frac{p-2}{p-1}\Big)^2\alpha^2
\epsilon_{e,-0.5}^2\epsilon_{B,-2}^{1/2}A_{\ast,-0.5}^{3/2}E_{53}^{-1}
\Gamma_{\rm min,1.8}^4 \: \rm Hz , \eeq

\beq
\nu_c\sim 1.9\times 10^{13}(1+z)^{-1}\alpha^2
\epsilon_{B,-2}^{-3/2}E_{53}A_{\ast,-0.5}^{-5/2}
\Gamma_{\rm min,1.8}^{-2} \: \rm Hz ,
\eeq

\beq F_{\nu,\rm max}\sim
385.0(1+z)D_{28}^{-2}\alpha^{-2}\epsilon_{B,-2}^{1/2}
A_{\ast,-0.5}^{3/2}\Gamma_{\rm min,1.8}^3\: \rm Jy , \eeq for the
wind case. It is shown in Table 1 that the temporal indices of
$\nu_m$, $\nu_c$ and $F_{\rm max}$ are the same as in the varying
injection model (Sari \& \Mesz 2000), which indicates that the
$\Gamma$-distribution in the shell does not affect the shape of
the light curve for the thin shell case (see Table 2) because the
spreading effect erases the initial $\Gamma$-distribution.

\section{NUMERICAL RESULTS}
Since the thin shell case in our treatment has the same
results as that
in Sari \& \Mesz (2000), hereafter we just present the numerical
results of the thick shell case. For the sake of simplicity, we
only show the result by one set of parameters with $s=2$ and
$b=2$, which is nevertheless sufficient to compare the result of
the thick shell case in our paper to that of the thin shell case
in the literature. We assume that the redshift of a GRB is $z=2$ since
it is the average value of the observed GRBs in the ${\textit
Swift}$ era (Le \& Dermer 2006). According to the standard shock
acceleration mechanism, the energy index $p$ of electrons is about
$2.2 \sim 2.3$, so we choose $p=2.3$.

We follow the method of Zou et al. (2005) to perform our numerical
calculations with $\Gamma_{\rm max}=1000$, $\Gamma_{\rm min}=300$,
$E_0=1.0\times 10^{53}$ erg, $\Delta_0=1.0\times 10^{13}$ cm,
$\epsilon_e=0.3$, $\epsilon_B=0.01$, $A_{\ast}=0.1$, $n_1=10~\rm
cm^{-3}$ and adopt the standard cosmology model with
$\Omega_m=0.27$, $\Omega_{\Lambda}=0.73$ and $H_0=71\rm \ km\
s^{-1}\ Mpc^{-1}$.

Figs.2 and 3 show the light curves of synchrotron emission in the
optical R band and X-ray band. To compare with the results
of the UEM,
we also plot the light curves of the UEM in the same energy band. In
our calculations, the nonuniform and uniform ejecta have the same
kinetic energy and total mass, so the uniform ejecta have a Lorentz
factor of $s\Gamma_{\rm min}/(s-1)=600$. The upper panel denotes the
R-band light curve while the lower panel denotes the X-ray light
curve. The numerical results demonstrate that: (1) In these two
energy bands, there is a slow decay phase before the crossing time
for the NUEM which is attributed to the energy injection by the
low-$\Gamma$ part. (2) Before the crossing time, the reverse shock
emission dominates the radiation in the optical band, while in the X-ray
band, the shocks have comparable contribution. After that the
forward shock emission gradually becomes important. This result is
similar to the UEM. It is possible that the two shocked regions may
have different microphysical parameters $\epsilon_e$ and
$\epsilon_B$, which do not change the first conclusion but might
change the second one.

Although we do not show the light curves of the thin shell case,
we would like to emphasize the differences between the thick and
thin shell cases. Since the reverse shock in the thin shell case
is mildly-relativistic all along and the flux density has the same
behavior as presented by the refreshed shock scenario (Sari \&
\Mesz 2000), a simple test of the thin shell case is that it
predicts a maximal flux in the far-infrared or millimeter range a
few hours to a few days after the GRB trigger. The forward shock
emission contributes mainly in the high energy band and decays slower
than that of the normal standard forward shock model. However, in
the thick shell case, the reverse shock may be relativistic and
dominate the flux in both the optical band and X-ray band at early times.

\section{DISCUSSION}
We have described the dynamics of radially structured ejecta
interacting with the circum-burst medium through extending the
method used in the UEM. It can be classified into two types: the thick
shell case and the thin shell case, of which the latter is the
same as in the UEM. Two parameters are introduced in the NUEM. One is the
mass distribution index $s$ and the other is the Lorentz factor
distribution index $b$. In our treatment, the thin shell case
reproduces the same results obtained by Sari \& \Mesz (2000). On
the other hand, in the thick shell case, the reverse shock could be
relativistic or initially be non-relativistic and then become
relativistic, contributing comparable radiation in the high energy
band as the forward shock. Anyway, the energy injection induced by
the nonuniform ejecta sweeping up the surrounding medium causes
the light curves to decay more slowly.

Observationally, GRB990123 was seen to have a bright optical
flash with initial flux decay as $F\propto t^{-2}$, which is
attributed to the reverse shock emission, and subsequently the
optical afterglow decays as $F\propto t^{-1.1}$, which is mainly
due to the forward shock emission. Both the thick and thin shell
scenarios in the UEM can fit this optical flash well. The fast rise of
$t^{3.4}$ can be explained  if the circum-burst environment is
homogenous ISM (Kobayashi 2000; Fan et al. 2002). However, some
optical flashes (e.g., GRBs 021211, 050525a, 060111B and 060117B)
have not been observed with this early rising part. This may be
intrinsic, or due to late responses and slow slewing of optical
telescopes. An early optical plateau was observed in a few GRB
afterglows, e.g., GRBs 050319, 060206, 060210 and 060313, which
may be attributed to the relativistic reverse shock emission of
the radially structured ejecta sweeping up the circum-burst
medium.

In $\textit{Swift}$ GRB afterglows, peculiar chromatic
breaks have been universally observed, but the origin of these
chromatic breaks is still an open question. Some models were
proposed to explain the chromatic breaks, such as the model with
evolving microphysical parameters, or the model in which the
optical and X-ray emission are arising from different emitting
regions (Panaitescu et al. 2006), or the scenario only involving
reverse shock emission (Genet et al. 2007; Uhm \& Beloborodov
2007). Our current work shows that the chromatic breaks
cannot be due to the reverse shock and we favor the former explanations.

Although the detailed prescription of a radially structured shell
propagating into the circum-burst medium is presented in this
paper, it should be noted that the shock-heated material separated
by the contact discontinuity is assumed to be uniform. A more
accurate solution of the reverse-forward shock interaction and
emission needs a relativistic hydrodynamic simulation.

\section{Acknowledgments}
This work was supported by the National Natural Science Foundation
of China (grants 10473023, 10503012, 10621303, 10633040, and
10703002), the National Basic Research Program of China (973
Program 2009CB824800). XFW also thanks the support of the NSF AST
0307376, NASA NNX07AJ62G, NNX08AL40G, the China Postdoctoral Science
Foundation, and the Postdoctoral Research Award of Jiangsu Province.

\label{lastpage}

\begin{table}
\begin{center}
\def\frh{}
\def\dn{2b(4-k)}
\def\dthin{2(7-2k+s)}
\begin{tabular}{cccccccccc}
\hline \hline
  & $\rm Reverse ~~Shock$   &$\rm Forward~~ Shock$ \cr
\hline
&\rm thick shell\hspace{1cm}thin shell &\rm thick shell\hspace{1cm}thin shell\cr

\hline
$\nu_m$ &$\frh\frac{5k-bk-ks-16}{\dn}$\hspace{1cm}
          -$\frh\frac{12-3k+ks}{\dthin}$
        &$\frh\frac{(s-1-3b)}{2b}$    \hspace{1cm}
          -$\frh\frac{24-7k+ks}{\dthin}$                 \cr

\hline
$\nu_c$ &$\frh\frac{(3k-4)(s+b-1)}{\dn}$\hspace{1cm}
         $\frh\frac{(3k-4)(s+1)}{\dthin}$
        &$\frh\frac{(3k-4)(s+b-1)}{\dn}$\hspace{1cm}
          $\frh\frac{(3k-4)(s+1)}{\dthin}$               \cr

\hline
$F_{\nu,\rm max}$    &  $\frh\frac{k+bk-3ks+10s-6b-2}{\dn}$ \hspace{0.5cm}
                     $\frh\frac{3(k+2s-4-ks)}{\dthin}$
                  &  $\frh\frac{3k-bk-3ks+8s-8}{\dn}$    \hspace{0.5cm}
                     $\frh\frac{k+6s-3ks-6}{\dthin}$           \cr

\hline
\hline
\end{tabular}
\caption{ Temporal indices of the peak frequency $\nu_m$, the
cooling frequency $\nu_c$ and the peak flux density $F_{\nu,\rm
max}$ for both the forward shock and reverse shock.}
\end{center}
\end{table}

\begin{table}
\begin{center}
\def\frh{}
\def\dn{2b(4-k)}
\begin{tabular}{cccccccccc}
\hline \hline & &$\rm thick\ shell\ case$  &\cr \hline &$\nu<
\rm{min}[\nu_m,\nu_c]$ &$\rm{min}[\nu_m,\nu_c]<\nu<
\rm{max}[\nu_m,\nu_c]$ &$\rm{max}[\nu_m,\nu_c]<\nu$ \cr

\hline
\rm Slow\ Cooling  &$\frac{5-9b-k+2bk+15s-4ks}{3b(4-k)}$
               &$\frac{12-12b-3k+3bk-16p+5kp-bkp+20s-5ks-kps}{4b(4-k)}$
               &$\frac{16-16b-6k+6bk-16p+5kp-bkp+16s-2ks-kps}{4b(4-k)}$\cr
\rm Fast\ Cooling  &$\frac{-5-7b+3k+17s-6ks}{3b(4-k)}$
               &$\frac{-16b-k+5bk+16s-3ks}{4b(4-k)}$
               &$\frac{16-16b-6k+6bk-16p+5kp-bkp+16s-2ks-kps}{4b(4-k)}$\cr

\hline \hline

& &$\rm thin\ shell\ case$ & \cr \hline

\rm Slow\ Cooling  &$\frac{12-3k-9s+4ks}{6k-3(7+s)}$
                   &$\frac{12-3k+12p-3kp-12s+5ks+kps}{8k-4(7+s)}$
                   &$\frac{16-6k+12p-3kp-8s+2ks+kps}{8k-4(7+s)}$\cr
\rm Fast\ Cooling  &$\frac{16-3k-11s+6ks}{6k-3(7+s)}$
                   &$\frac{28-9k-8s+3ks}{8k-4(7+s)}$
                   &$\frac{16-6k+12p-3kp-8s+2ks+kps}{8k-4(7+s)}$\cr

\hline
\hline
\end{tabular}
\caption{Temporal indices of the flux density $F_\nu\propto
t^{\alpha}\nu^{\beta}$ of synchrotron radiation from a reverse
shock. Both the thick shell
  case and the thin shell case are considered.}
\end{center}
\end{table}

\begin{figure}
\centering
\includegraphics[angle=360,width=1.0\textwidth]{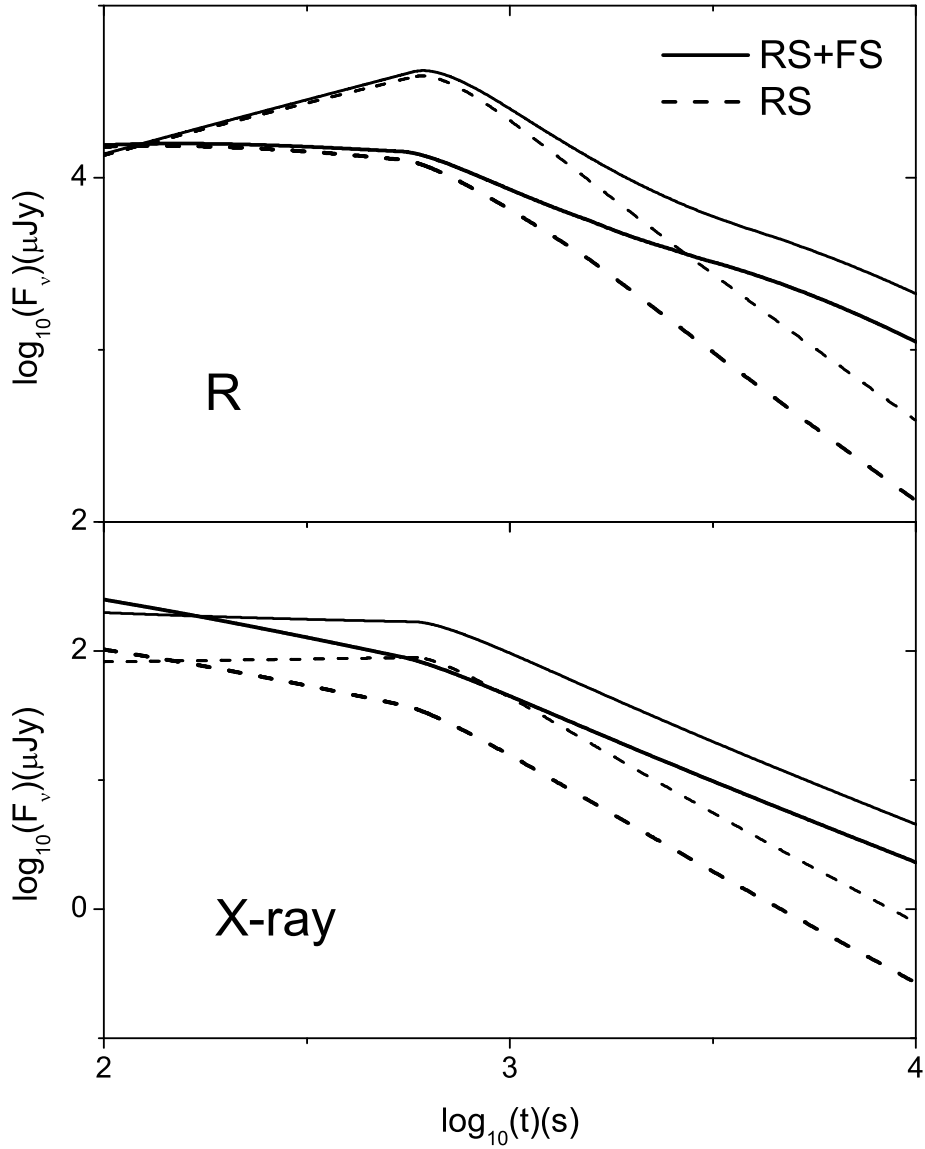}
\caption{Synchrotron radiation flux density in the R band and X-ray band as
  a function of time for the ISM case. Solid
  lines represent the whole emission from both the forward
  shock and the reverse shock. Dashed lines represent the
  contribution from the reverse shock. Thick lines are for
  the NUEM while thin lines are for the UEM.}
\label{fig2}
\end{figure}

\begin{figure}
\centering
\includegraphics[angle=360,width=1.0\textwidth]{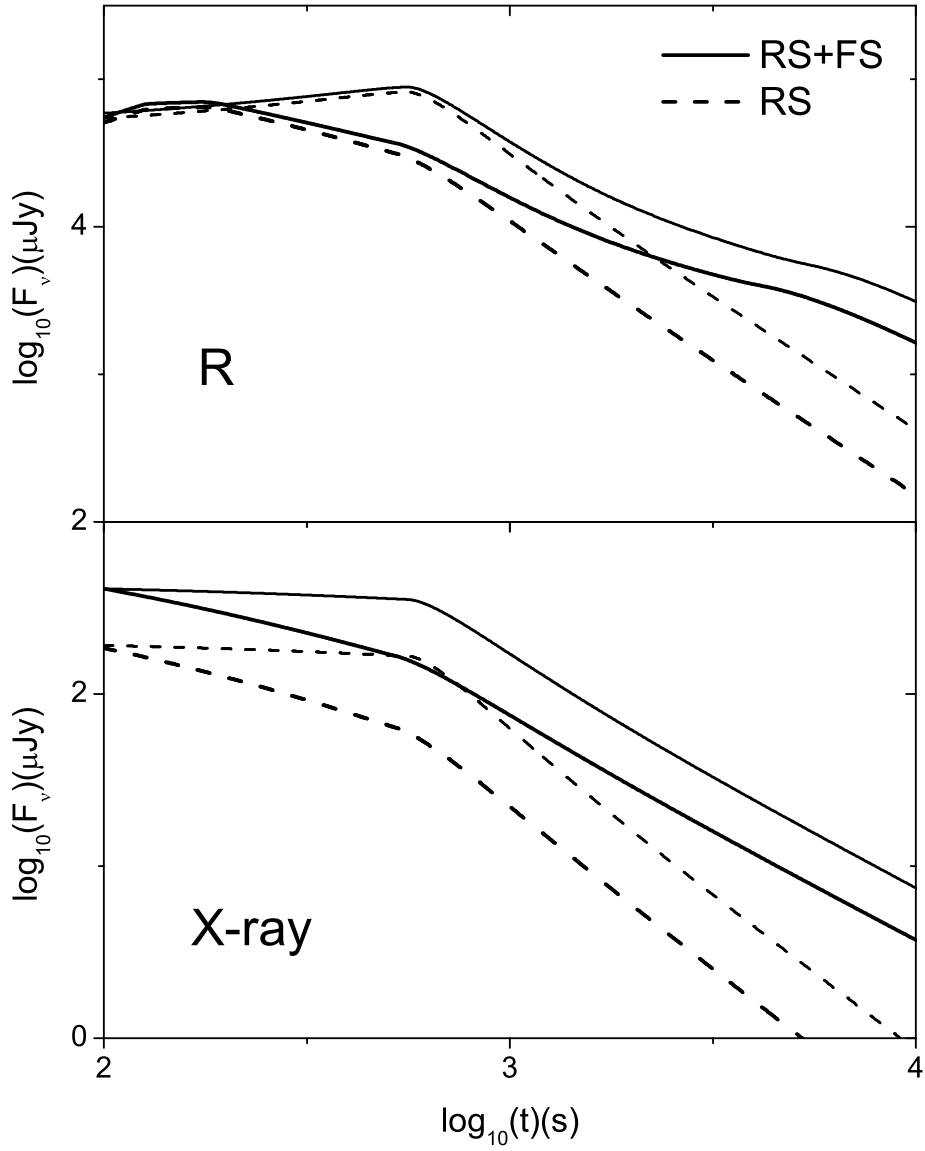}
\caption{Synchrotron radiation flux density in the R band and
  X-ray band as a
  function of time, the same as in Fig. 2. for the wind case.
  Parameters are given in the text.}
\label{fig3}
\end{figure}


\begin{thebibliography}{99}
\bibitem[Akerlof et al.(1999)]{Akerlof99}Akerlof C. et al., 1999, Nature,
  398, 400
\bibitem[BM76]{BM76}Blandford R. D., McKee, C. F., 1976, Phys. Fluids,
  19, 1130
\bibitem[Chevalier \& Li(2000)]{Chevalier99} Chevalier R. A.,
  Li Z. Y., 2000, ApJ, 536, 195
\bibitem[Costa \etal(1997)]{Costa97}Costa E., et al., 1997, Nature,
  387, 783
\bibitem[Fan \etal(2002)]{Fan02} Fan Y. Z., Dai Z. G., Huang, Y. F., Lu, T., 2002, ChJAA, 2, 449
\bibitem[Genet \,Daigne \& Mochkovitch(2007)]{Genet07} Genet
  F., Daigne F., Mochkovitch R., 2007, MNRAS, 381, 732
\bibitem[Jelinek \etal(2006)]{Jelinek06} Jelinek M. et al., 2006,
  A\&A, 454, L119
\bibitem[Klotz \etal(2006)]{Klotz06} Klotz A. et al., 2006, A\&A,
  451, L39
\bibitem[Kobayashi \etal(1999)]{Kobayashi99}Kobayashi S.,
  Piran T., Sari R., 1999, ApJ, 513, 669
\bibitem[Kobayashi(2000)]{Kobayashi00} Kobayashi, S., 2000,
  ApJ, 545, 807
\bibitem[Le \& Dermer(2006)]{Le06}Le T., Dermer C. D., 2007,
  ApJ, 661, 394
\bibitem[\Mesz \& Rees 1993]{Mes93}\Mesz P., Rees M. J., 1993, ApJ, 405, 278
\bibitem[\Mesz \& Rees 1997]{Mes97}\Mesz P., Rees M. J., 1997, ApJ, 476, 232
\bibitem[Molinari \etal(2007)]{Molinari07}Molinari E. et
  al., 2007, A\&A, 469, L13
\bibitem[Nousek (2006)]{Nousek06}Nousek J. A. et al., 2006, ApJ, 642,389
\bibitem[O'Brien \etal (2006)]{Brien06}O'Brien P. T. et al., 2006, ApJ, 647, 1213
\bibitem[Panaitescu \etal (2006)]{Panai06}Panaitescu A., \Mesz
  P., Burrows, D. et al., 2006, MNRAS, 369, 2059
\bibitem[Racusin \etal 2008]{Racusin08} Racusin J. L. et al., 2008, Nature, 455,
183
\bibitem[Rees \& \Mesz 1992]{Rees92} Rees M. J., \Mesz P. 1992, MNRAS, 258, 41
\bibitem[Rees \& \Mesz 1994]{Rees94} Rees M. J.,\Mesz P., 1994, ApJ, 430, L93
\bibitem[Rees \& \Mesz 1998]{Rees98} Rees M. J.,\Mesz P., 1998, ApJ, 496, L1
\bibitem[Sari \& Piran (1999)]{Sari99} Sari R., Piran T., 1999, ApJ,517, L109
\bibitem[Sari \& \Mesz 2000]{Sari02} Sari R.,\Mesz P., 2000, ApJ, 535, L33
\bibitem[Sari \& Piran 1995]{Sari95} Sari R., Piran T.,
  1995, ApJ, 455, L143
\bibitem[Shao 2005]{Shao05}Shao L.,Dai Z. G., 2005, ApJ, 633, 1027
\bibitem[Uhm \& Beloborodov 2007]{Uhm07}Uhm Z. L.,
  Beloborodov A. M., 2007, ApJ, 665, L93
\bibitem[Wei 2003]{Wei03}Wei D.M., 2003, A\&A, 400, 415
\bibitem[Wu \etal 2003]{Wu03} Wu X. F., Dai, Z. G., Huang,
  Y. F., Lu T., 2003, MNRAS, 342, 1131
\bibitem[rongrong]{xue09}Xue R. R., Fan Y. Z., Wei D. M.,
  2009, A\&A, 498, 671
\bibitem[Zhang \etal 2003]{Zhang03} Zhang B., Kobayashi, S., \Mesz P., 2003, ApJ, 595, 950
\bibitem[Zhang \etal 2006]{Zhang06} Zhang B., Fan Y, Z., Dyks J. et al., 2006, ApJ, 642, 354
\bibitem[Zhang 2007]{Zhang07} Zhang B., 2007, ChJAA, 7, 1
\bibitem[Zou \etal 2005]{Zou05} Zou Y. C., Wu X. F., Dai Z. G., 2005, MNRAS, 363, 93

\end{thebibliography}
\end{document}